\documentclass[aps,prl,twocolumn,superscriptaddress]{revtex4}
\usepackage{longtable}
\usepackage[dvips]{graphicx}
\def\be{\begin{equation}}
\def\ee{\end{equation}}
\def\bea{\begin{eqnarray}}
\def\eea{\end{eqnarray}}
\def\ba{\begin{array}}
\def\ea{\end{array}}
\def\abs#1{\vert #1 \vert}

\begin{document}

% Use the \preprint command to place your local institutional report
% number in the upper righthand corner of the title page in preprint mode.
% Multiple \preprint commands are allowed.
% Use the 'preprintnumbers' class option to override journal defaults
% to display numbers if necessary
%\preprint{}

%Title of paper
\title{A B-spline Galerkin method for the Dirac equation }

% repeat the \author .. \affiliation  etc. as needed
% \email, \thanks, \homepage, \altaffiliation all apply to the current
% author. Explanatory text should go in the []'s, actual e-mail
% address or url should go in the {}'s for \email and \homepage.
% Please use the appropriate macro foreach each type of information

% \affiliation command applies to all authors since the last
% \affiliation command. The \affiliation command should follow the
% other information
% \affiliation can be followed by \email, \homepage, \thanks as well.
\author{Charlotte Froese Fischer}
\email{Charlotte.Fischer@nist.gov}
%\homepage[]{Your web page}
%\thanks{}
%\altaffiliation{}
\affiliation{Atomic Physics Division, National Institute of Standards and
Technology, Gaithersburg, Maryland 20899-8422}
\author{Oleg Zatsarinny}
\affiliation{Department of Physics and Astronomy, Drake University, Des Moines, IA 50311, USA}

%Collaboration name if desired (requires use of superscriptaddress
%option in \documentclass). \noaffiliation is required (may also be
%used with the \author command).
%\collaboration can be followed by \email, \homepage, \thanks as well.
%\collaboration{}
%\noaffiliation

\date{\today}

\begin{abstract}
The B-spline Galerkin method is investigated for the
simple eigenvalue problem, $y^{\prime\prime} = -\lambda^2 y$. 
%that can also be.expressed as a pair of first-order differential equations.
%We show that solutions of the latter are the same as the former
%when a $(B,B')$ basis is used which, in turn, is equivalent to a 
%$(B^k,B^{k^{-1}})$ basis.  
Special attention is give to boundary conditions.
From this analysis, we
propose a stable method for the Dirac equation and evaluate 
its accuracy by comparing the computed and exact R-matrix for a wide range
of nuclear charges $Z$ and angular quantum numbers $\kappa$.  
No spurious solutions were found and excellent agreement was obtained
for the R-matrix. 

\end{abstract}

%\maketitle must follow title, authors, abstract, \pacs, and \keywords
\maketitle

%\end{widetext}

%\tableofcontents

%\listoftables

% body of paper here - Use proper section commands
% References should be done using the \cite, \ref, and \label commands
% Put \label in argument of \section for cross-referencing
%\section{\label{}}
%\subsection{}
%\subsubsection{}

% \section{Introduction}

The B-spline methods Johnson and Sapirstein ~\cite{johnson,sapirstein} introduced into
relativistic many-body perturbation theory have produced results of unprecedented
accuracy~\cite{WRJ08}. Essentially, the local non-orthogonal B-spline basis was
transformed to an orthogonal orbital basis by the application of the Galerkin method to
the Dirac equation over a finite interval~\cite{WRJ88}. The resulting
basis was  finite and effectively complete. Though the low-energy bound states were good
approximations to solutions of the Dirac equation, no physical interpretation was
important for continuum states. Rapidly oscillating solutions were observed
but played a negligible role in the summation over states in their 
applications~\cite{sapirstein}.
However, these spurious states perturbed the spectrum and slowed
the convergence of quantum electrodynamic (QED) calculations. 
This led 
Shabaev {\it et al.}~\cite{shabaev} to propose a dual kinetic balance basis 
similar to the basis
Quiney {\it et al.}~\cite{kinetic} employed with analytic Slater type functions.
Boundary conditions were for the case of a finite nuclear-charge
distribution, with the point nucleus considered as a limiting case. Different
boundary conditions at the origin were proposed for positive and negative
values of $\kappa$ and both large and small components were set to zero at
the large $r$ boundary.
Recently Igarashi~\cite{japan} investigated a variety of methods and boundary 
conditions.  He
pointed out that the four boundary conditions used by Froese Fischer and
Parpia~\cite{cffP} were excessive and explored the use of B-splines of 
different order,
$k_p$ and $k_q$, as a way of avoiding spurious solutions.
% observing that the method worked best when $| k_p - k_q | =1$.
In a subsequent paper he  concluded that kinetic balance also provided a 
good basis ~\cite{japan2}. No best method was identified.  All his methods
employed analytic weighting factors to B-spline expansions in order
to control the asymptotic properties of large and small components.

R-matrix methods (see Ref.'s~\cite{grant,dbsr} for recent reviews) differ from
the applications considered by the above authors in that zero boundary 
conditions at large $r$,
such as proposed by Shabaev {\it et al.}~\cite{shabaev}, cannot be used.
R-matrix theory assumes an inner region $r < a$
in which exchange is important and an outer region $r >a $ 
where exchange with an outer electron can be neglected. What is needed is a 
basis for the inner region that satisfies certain conditions at the $r=a$ 
boundary.  B-splines were very successfully employed in the non-relativistic 
R-matrix calculations~\cite{bsr}, however, they cannot be used in the 
Dirac-based calculations when spurious states in the continuum spectrum are 
present. At the same time, the kinetically balanced bases lead to extensive
computational difficultes in many-electron calculations.

Spline methods are based on approximation theory. The grid that is selected
along with boundary conditions 
determine a piecewise polynomial space with a finite basis.
% ${\cal P}_{k,\xi}$ where $k$ denotes
%the order and $\xi$ the grid, 
%which we assume is one-dimensional.  
The unique B-spline basis has many advantages ~\cite{bachau,bsr}, 
but there are many possible bases. 
 The transformation from a non-orthogonal 
basis to an orthogonal orbital basis
depends on how the Galerkin method is applied.
In this letter we propose a simple method for the Dirac
matrix equation and apply it to the calculation
of the R-matrix boundary condition. All calculations are for a point 
nucleus so that results can be compared with exact solutions. 
 Special attention is given to the
boundary conditions.  We also show the relationship between
kinetic balance and the use of splines of different order.

At large values of $r$, the non-relativistic Schr\"odinger equation has the
same form as
\be
y^{\prime\prime}(r) = -\lambda^2 y(r), \; y(0) = 0,
\ee
for which the solutions are $y(r) = c \sin(\lambda r)$.
 A second boundary condition at  $r = a$ defines
the allowed values of $\lambda$.  With $y(a)=0$, the values of $\lambda$ are
such that $\lambda a = n\pi$ and $n$ an integer. We will denote the computed
value of $n$ as $n^*$. 
%or $\lambda^2 = (n\pi/a)^2$.

\begin{figure}[t]
%\vspace{24pt}
\flushleft{ \includegraphics[width=3.0in]{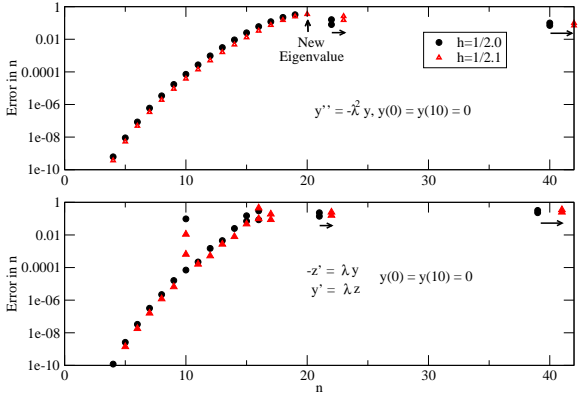}}
\caption{ Errors in $n$ for the Galerkin method
applied to equivalent systems of differential equations.}
\label{fig1}
\end{figure}

With a grid consisting of subintervals of length $h$ and knots of 
multiplicity $k$ at $r=0$ and $r=a$,
 the solution  $y(r) =
\sum_2^{N-1}y_iB_i(r)$  satisfies $y(0)= y(a) = 0$, where $N$ is the size of the
basis. The Galerkin requirement that the  residual be
orthogonal to each basis element in the expansion, leads to the generalized
matrix eigenvalue problem \be
 D^{(02)}\, y = -\lambda^2  B \, y
\ee where  $D^{(02)}(i,j) = \langle B_i(r) | B^{\prime\prime}_j(r) \rangle$ and $B(i,j) =
\langle B_i(r) | B_j(r) \rangle$. The superscripts on the the derivative matrix designate
the order of the derivatives acting on $B_i(r)$ and $B_j(r)$, respectively. 
The top graph of Fig.~\ref{fig1}
shows the error $\abs{n-n^*}$
 as a function of $n$  for two different grids that increase the matrix size
by one.  Splines of order 6 were used.
The steady reduction in accuracy as $n$ increases is expected but
not all solutions are approximations to solutions of the differential equation
for which a notion of convergence should apply. As $h$ is reduced 
and the size of the matrix increased by one,
a new eigenvalue appears but the four highest move to higher values.
The latter four are not approximations to the differential equation
but are needed for the completeness of the orthogonal basis set. 

\begin{figure}
%\vspace{24pt} 
\flushleft{ \includegraphics[width=3.0in]{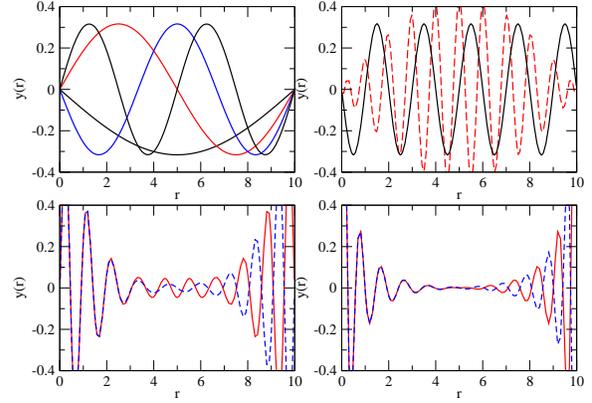}} \caption{ Some
solutions of $-z^\prime(r)=ky(r),y^\prime(r) = kz(r)$
 with $y(0)=y(10)=0$: top left, $n=1-4$; top right, $n=10,10$;
bottom left, $n=21,21$; bottom right, $n=39,39$ (see Fig. \ref{fig1}). }
 \label{fig2}
\end{figure}

The second-order differential equation can also be written as a pair of
first-order equations in a form similar to that of the Dirac equation,
namely
\bea
 \left[ \begin{array}{c c }
     0  & -d/dr  \\
     d/dr  &  0 \\
    \end{array} \right] \left[ \begin{array}{c}
                                       y(r) \\
                                       z(r) \\
                                \end{array}\right] = k \left[ \begin{array}{c}
                                                           y(r) \\
                                                           z(r) \\
                                                       \end{array} \right] .
\eea
The solution of these equations with $y(0) = 0$ are $y(r) = c \sin(kr)$ and
$z(r) = c \cos(kr)$.  Note that the boundary condition $y(0) = 0$ implies
$z(0) =c$ and $y(a)=0$ implies $z(a) = \pm c$.  Thus no boundary condition
should be applied on the latter.  With the assumption
\be
y(r) = \sum_{i=2}^{N-1} y_i B_i(r) \mbox{\ and\ }
z(r) = \sum_{i=1}^{N} z_i B_i(r),
\ee
the Galerkin method leads to the generalized eigenvalue problem
\bea
\label{eq-M1A}
 \left[ \begin{array}{c c }
     0  & D^{(10)}  \\
     D^{(01)}  &  0 \\
    \end{array} \right] \left[ \begin{array}{c}
                                       y \\
                                       z \\
                                \end{array}\right] = k
                        \left[ \begin{array}{c c}
                                B & 0 \\
                                0 & B \\
                               \end{array} \right] \left[ \begin{array}{c}
                                                           y \\
                                                           z \\
                                                          \end{array}\right].
\eea
with $D^{(01)}$ a rectangular matrix of N rows and N-2 columns and 
$D^{(10)}$ its transpose.
The values of $\lambda$ occur in positive and negative pairs except for two eigenvalues with
$\lambda=0$, a consequence of $D^{(01)}$ being rectangular. 
The bottom graph of Fig. \ref{fig1} shows the errors in $n$ for
 the positive eigenvalues.
Note the presence of two solutions for $n= 10,15,16$ and higher.
The error of only one of the two solutions decreased as the 
step-size was made smaller.
Fig. \ref{fig2} shows some of the eigenfunctions.  Solutions of the differential
equation have constant amplitude as seen for $n=1-4$.  Of the two $n=10$ 
solutions, one has constant amplitude whereas the other does not
and is referred to as 
a spurious solution. The four eigenfunctions of the bottom graphs
are ``border'' solutions and correspond to solutions in 
Fig.~\ref{fig1} that move to higher energies as $h$ is decreased.
The numerical properties of the solutions of the two 
equivalent differential equations can be interpreted by comparing
the errors depicted in Fig.~\ref{fig1} 
as $h$ is decreased and
the matrix size increases by one. In the top graph,
 the errors of all eigensolutions decrease, a new solution 
appears, and border solutions move to higher energies.
In the bottom graph some 
errors decrease, the new eigensolution may be a spurious solution, and
border solutions move to higher energy.
Clearly the second-order differential equation solution whose errors 
are depicted in the top graph is the preferred solution.

Unlike the differential equations where $z(r)$ can be eliminated to yield the original
second-order differential equation, the elimination of the vector $z$ leads to 
the matrix
$D^{(10)} B^{-1}D^{(01)}$ for $y$ that is quite different from $D^{(02)}$.  
Whereas the latter matrix for the B-spline basis is banded, the former is a 
full matrix.
The difference is most dramatic if the matrices are evaluated in the basis of
eigenvectors of $D^{(02)}$.  Then $D^{(02)}$ is diagonal whereas 
$D^{(10)} B^{-1}D^{(01)}$
is striated with a non-zero diagonal and alternating
 zero and non-zero sub- or super-diagonals. 

The pair of first-order equations could also be solved by assuming
$$y(r) = \sum_{i=2}^{N-1} y_i B_i(r) \mbox{\ and\ } z(r) = \sum_{i=2}^{N-1} z_i B^{\prime}(r).$$
 Then the application of the Galerkin method leads to
\bea
 \left[ \begin{array}{c c }
     0  &   D^{(11)}  \\
     D^{(11)}  &  0 \\
    \end{array} \right] \left[ \begin{array}{c}
                                       y \\
                                       z \\
                                \end{array}\right] = k
                        \left[ \begin{array}{c c}
                                B & 0 \\
                                0 & D^{(11)} \\
                               \end{array} \right] \left[ \begin{array}{c}
                                                           y \\
                                                           z \\
                                                          \end{array}\right].
\eea
Furthermore, $D^{(11)}$ is symmetric and positive definite.
Thus the second set of equations
defines the relationship between the expansion coefficients, namely $y_i = kz_i$
and the equation for $y$ is the same as for $y^{\prime\prime}
 = -\lambda^2 y$ since, with our boundary
conditions,  $D^{(02)} = -D^{(11)}$. The numerical results are identical.
 
Each basis function, $B^\prime_i(r), i=2,\ldots, N$ is a spline of order $k-1$.
This set defines a complete, linearly independent basis for a
spline approximation of order $k-1$ (for simplicity, the superscript $k$ is
omitted for splines of order $k$).  Another basis for splines of order $k-1$
are the B-splines, $B_i^{k-1}(r), i=1,\ldots,N-1$. In fact, the first derivative
of a spline function can be found by differencing its B-spline 
coefficients~\cite{deboor}. Thus this method is
similar to the method proposed by Igarashi~\cite{japan} provided
$| k_p -k_q | =1$ and without analytic weighting factors. An
expansion in the ($B,B^\prime$) basis is also 
similar to a kinetically balanced 
basis with $\kappa = 0$ and is
equivalent to an expansion in the ($B^k, B^{k-1}$) basis. 
The advantage of the latter is
that all basis functions are again strictly positive functions and boundary
conditions are simpler to apply.

Some further comments are in order.  If in Eq. (2), the matrix $D^{(02)}$ is
replaced by $-D^{11}$, no boundary condition is needed to preserve
symmetry at $r=a$.  These two
matrices differ only in the last column and the difference can be treated
as a symmetrizing Bloch operator~\cite{bsr}.
 The resulting spectrum is for $n= 1/2, 3/2, \ldots$ for which
$y^\prime(a)=0$.
 The numerical accuracy 
of the modified Eq.(2) and extended Eq. (6) is unchanged.

%In this letter we propose a general, stable approach 
%for the solution of the Dirac
%equation that is based on two sets of B-splines and we describe its 
Based on the above findings for B-spline solutions of
differential equations we propose a general, stable method
for the Dirac equation and describe its
application to the R-matrix method.
For a single electron in the Coulomb potential,
$V(r) = -Z/r$, and a point nucleus of charge $Z$ the equation
   may be written as
\bea
\label{eq:DC}
\left[ \begin{array}{c c}
      V(r) & -c\left[ \frac{d}{dr} -\frac{\kappa}{r}\right] \\
      \ \\
     c\left[ \frac{d}{dr} +\frac{\kappa}{r}\right] & V(r) -2c^2
     \end{array}\right] \left[\begin{array}{c}
      P(r)\\
      Q(r)
     \end{array}\right] = E \left[\begin{array}{c}
      P(r)\\
      Q(r)
     \end{array}\right].
\eea

The R-matrix method requires an effectively complete
basis $(P_i,Q_i)$ for the inner $r < a$ region with
$P_i(0) = 0$ and 
special boundary conditions at $r=a$ that determine the set of
energies $E_i$.  For low positive energies $E$, and $r=a$
sufficiently large so that $V(a)$ is small relative to $-2c^2$,
 it follows that $Q(a)\approx 
(a P^\prime(a) + \kappa P(a))/2ac$~\cite{NG}.
The boundary conditions for the desired
 R-matrix solutions of the Dirac equation
are $Q_i(a)/P_i(a) = (b+\kappa)/(2ac) = p$, where $b$ is an 
arbitrary constant.  Thus both large and small components
are non-zero on the boundary though not equal.

With non-zero solutions on the boundary, the Galerkin method
as described earlier will not yield a symmetric matrix.  Variational
methods applied to an associated action (\cite{WRJ88}, Eq. 7)
can be used, but in R-matrix theory it is customary to apply 
a Bloch operator that enforces the boundary condition as 
well as symmetry. This operator is then also used for the outer
region.   Let
\be
\hat{\cal H} = {\cal H} + {\cal L}
\ee
where ${\cal H}$ is the Dirac operator of Eq. (\ref{eq:DC}) and ${\cal L}$
is the Bloch operator~\cite{SH}
 \bea \label{eq:Bloch}
 {\cal L} = c\delta(r-a)\left( \begin{array}{c c }
     -p\eta & \eta  \\
     (\eta-1) &  (1-\eta)/p  \\
    \end{array} \right),
\eea and $\eta$ is an arbitrary constant. In the
present calculations we have used the values $\eta=0.5$ and
$p=\kappa/2ac$.  

Having defined the operators, let us now define the spline expansions.
Suppose there are two sets of B-splines on the same grid that define
the $(B^{k_p}, B^{k_q})$ basis for the Dirac equation.
Then the number of functions in each set are $n_p=n_v+k_p-1$ and $ n_q=n_v+
k_q-1$, respectively where $n_v$ is the number of intervals.
The expansions
$P(r)= \sum_{i=2}^{n_p} p_i B_i(k_p;r)$ 
$Q(r) = \sum_{i=1}^{i=n_q} q_i B_i(k_q;r)$
satisfy the boundary condition $P(0)=0$ and
lead to 
\bea \label{eq:BD}
 \left[ \begin{array}{c c }
     V_{11}  & W_{12}  \\
     W_{21}  & V_{22}  \\
    \end{array} \right] \left[ \begin{array}{c}
                                       p \\
                                       q \\
                                \end{array}\right] = E
                        \left[ \begin{array}{c c}
                                B_{11} & 0 \\
                                0 & B_{22} \\
                               \end{array} \right] \left[ \begin{array}{c}
                                                           p \\
                                                           q \\
                                                          \end{array}\right],
\eea
where
\bea
V_{11}(i,j) & = & \langle B_i^{k_p}|-Z/r|B_j^{k_p} \rangle 
         -(cp/2)\delta_{in_p}\delta_{jn_p} \nonumber \\
V_{22}(i,j) & = & \langle B_i^{k_q}|-Z/r -2c^2|B_j^{k_q} \rangle
          +(c/2p)\delta_{in_q}\delta_{jn_q}
            \nonumber \\
W_{12}(i,j) & = &-c\langle B_i^{k_p}| \frac{d}{dr} -\frac{\kappa}{r}|
                                             B_j^{k_q}\rangle 
        +(c/2)\delta_{in_p}\delta_{jn_q}\nonumber \\
W_{21}(j,i) & = &c\langle B_j^{k_q}| \frac{d}{dr} +\frac{\kappa}{r}
                          | B_i^{k_p}\rangle
                 -(c/2) \delta_{in_q} \delta_{jn_p}, 
\eea and $B_{11}$ and $B_{22}$ are the overlap matrices for $B^{k_p}$ 
and $B^{k_q}$,
respectively.  We have used the fact that for a grid 
with muliple knots at $r=a$, $B_N(a) =1$ for all orders.

From the above finite set of solutions, 
an R-matrix relation can be derived that connects the
inner and outer region.
For a given energy $E$, the relation has the form
\be
 \label{eq:PQ}
  P(a)=\left[ R(E)-\frac{b+\kappa}{(b+\kappa)^2+(2ac)^2} \right][2acQ(a)-(b+\kappa)P(a)]
\ee
where the relativistic $R$-matrix is defined as
\be
 \label{eq:RM}
  R(E)=\frac{1}{2a}\sum_{i}\frac{P_i(a)P_i(a)}{E_i-E}.
\ee Eq. (\ref{eq:PQ}) contains the correction 
$(b+\kappa)[(b+\kappa)^2+(2ac)^2]$,
first obtained by Szmytkowski and Hinze ~\cite{SH}. This correction
is due to the fact that
the set of relativistic basis functions $(P_i,Q_i)$ is incomplete
 on the surface $r=a$.
However, it is small in most
realistic cases and usually is omitted.

\begin{figure}
\centerline{ \includegraphics[width=3.0in]{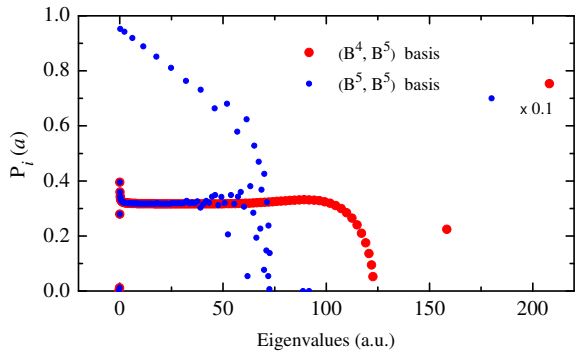}} 
\caption{Comparison of the surface
amplitudes for $(B^4,B^5)$ and $(B^5,B^5)$ bases in the case $Z$=1,
$\kappa=-1$, $a=20$, $N=100$, on an equally spaced grid.} 
\label{amp}
\end{figure}

If we employ expansions with $k_q=k_p$ in Eq. (\ref{eq:BD}),
many pseudo-solutions are found in the
positive-energy spectrum. These
pseudo-solutions are characterized by a rapidly oscillating behavior
with every coefficient in the B-spline expansion changing sign: they
cannot be used directly in Eq. (\ref{eq:RM}) for the R-matrix.
The use of B-splines of different order removes all
pseudo-solutions. The $(B^k,B^{k+1})$ basis was found to
be the most stable numerically. This stability is extremely important
for the calculation of $R(E)$.
Fig.~\ref{amp} compares the surface amplitudes $P_i(a)$.
%, which are the main parameters for the R-matrix definition.
In the case of the $(B^4,B^5)$ basis the surface amplitudes vary smoothly with
energy, whereas the $(B^5,B^5)$ basis produces many pseudo-solutions with large
surface amplitudes. The surface amplitudes for these two calculations agree 
only for some
low energy eigenstates, but differ considerably in the higher energy spectrum.
 The high
energy eigensolutions in both cases are not pseudo-solutions, 
though they have very
large surface amplitudes. As in the model equation, these border solutions 
are needed for
the effective completeness in the transformation from a B-spline basis to 
eigenstates of
the Dirac matrix equation. These eigenstates provide a relatively large contribution to
the total value of $R(E)$ that brings the final value in closer agreement with the
exact value. Note that Fig.~\ref{amp} shows only positive-energy (electron) solutions.
Contributions to the R-matrix (\ref{eq:RM}) from the negative-energy (positron) solutions
were found to be negligibly small in the present case.

\begin{figure}
\centerline{ \includegraphics[width=3.0in]{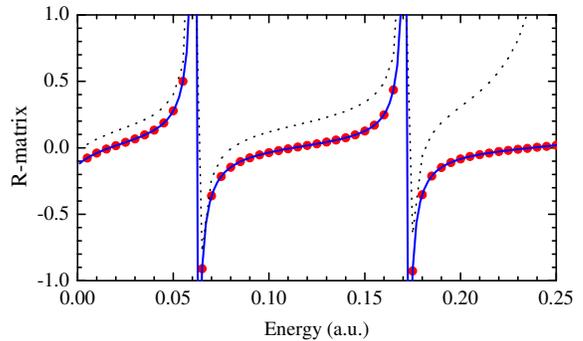}} \caption{Comparison of the exact
R-matrix (blue line) with B-spline R-matrix in the $(B^4,B^5)$ basis (red circles) and in
the $(B^5,B^5)$ basis (dotted line) for the same case as Fig. 3.} 
\label{RM}
\end{figure}

In the case of the Dirac-Coulomb problem with a point nucleus,
we can check directly the accuracy of the resulting
R-matrix because the wavefunctions are known analytically.
% In the case of
%the boundary conditions (\ref{eq:BD})
The R-matrix  can be expressed as \be
 \label{eq:RMC}
  R(E)=\left[2acG\left(E\right)-(b+\kappa)F\left(E\right)\right]/F(E),
\ee where $F(E)$ and $G(E)$ are the large and small components of the Dirac-Coulomb
wavefunction for given $\kappa$ and $Z$. Comparison of the exact R-matrix 
with the
one obtained from B-spline bases is shown in Fig.~\ref{RM}. There is very close
agreement with the exact results for a wide range of energy for the $(B^4,B^5)$
basis (only the low-energy region is shown for better visualization).  The 
results correctly reproduce the tangent-like behavior of the
exact R-matrix, along with a
correct representation of all poles. At the same time, the $(B^5,B^5)$ basis
lead to large errors due to the presence of pseudo-states.

The ($B^4, B^5$) method with $a=20/Z$ and $N=100$ was checked for
a wide range of $Z$ and with $\kappa$ up to $\pm 50$. No spurious solutions 
were 
found. The accuracy of the R-matrix for small $Z$ and all $\kappa$ was in the
range $10^{-6}$ to $10^{-8}$, decreasing for large $Z$.  At 
$Z=100$ the accuracy had deteriorated to $10^{-3}$ but no attempt was made to
modify the grid or change ($k_p$, $k_q$). 
 The R-matrix was relatively independent of
whether an equally spaced or exponential grid was used although the behavior
near the origin was not monitored.

We have also checked the accuracy of the R-matrix calculations for the 
kinetic balance
B-spline basis proposed by Igarashi ~\cite{japan2} but
omitting analytic factors. The resulting accuracy is
approximately the same as for $(B^{k},B^{k+1})$ or $(B,B')$ bases but the 
method is much more difficult to implement, 
especially in the case of multi-channel R-matrix calculations,
since different bases are needed for different values of $\kappa$.
The dual kinetic balance basis, proposed by Shabaev et al.~\cite{shabaev} 
failed to reproduce an accurate R-matrix, because it resulted in many 
pseudo-solutions in the
non-physical energy region just above the $-mc^2$ threshold.
We also found that the appearance of pseudo-solutions depends only in a minor 
way on initial or boundary conditions. In fact the most accurate results are 
obtained with a minimum of
additional conditions on the B-spline coefficients.
% Calculations have also

In conclusion, a simple but stable method is proposed for the solution of 
the Dirac
equation, including the eigenvalue problems arising in R-matrix theory. 
Whereas earlier considerations concentrated on the non-relativistic limit
of the Dirac equation,
we have shown the importance of the large $r$ region.  Any reliable
method for the Dirac equation
must be able to solve the pair of first-order equations of Eq.(3)
to the same accuracy as Eq.(2) or, equivalently have matrices for which
$D^{(02)} = D^{(10)} B^{-1}D^{(01)}$. 
 In general, accurate methods 
require an exponential grid near the origin
 in order to reproduce the
$r^{\gamma}$ behavior where $\gamma = \sqrt{\kappa^2
- Z^2/c^2}$ but the singularity at the origin itself has not been found to 
be a problem. 

The methods described here have been applied to the investigation of low-energy
electron scattering from Cs~\cite{dbsr}. A finite nucleus was used along with 
B-splines of order (8,9). Close agreement with experiment was obtained for
the total and angle differential cross sections as well as several 
spin-asymmetry parameters.

%The latter are computationally much more  effective because they are not
%dependent on the orbital parameter $\kappa$, a factor that is
%extremely important in many-body calculations. 
%We show that the problem
%of pseudo-solutions is a common problem for a wide class of system of differential
%equations with hidden symmetries.
%The bases which brake these symmetries are more stable
%against the appearance of pseudo solutions.
%.been checked for high values of $Z$ and $\kappa$ where an exponential grid 
%is needed. No problems were encountered.  In particular, no additional boundary
%conditions at the origin were needed to eliminate spurious solutions.

%\section*{Acknowledgments}

The work of O. Z.  was supported by the National Science Foundation under 
Grant No.  PHY-0244470.

%\section{References}


\begin{thebibliography}{10}

\bibitem{johnson} W. R. Johnson and J. Sapirstein, Phys. Rev. Lett.
{\bf 57}, 1126 (1986).

\bibitem{sapirstein} J. Sapirstein, and W. R. Johnson, J. Phys. B,
{\bf 29}, 5213 (1996).

\bibitem{WRJ08} W. R. Johnson, U. I. Safronova, A. Derevianko, M. S.
Safronova, Phys. Rev. A {\bf 77}, 022510 (2008).

\bibitem{WRJ88} W. R. Johnson, S. A. Blundell, and J. Sapirstein,
Phys. Rev. A {\bf 37}, 307 (1988).

\bibitem{shabaev} V. M. Shabaev, I. I. Tupitsyn, V. A. Yerokhin, G. Plunien,
G. Soff, Phys. Rev. Lett. {\bf 93}, 130405 (2004).

\bibitem{kinetic} H. M. Quiney, I. P. Grant, and S. Wilson, Phys. Scripta
{\bf 36}, 460 (1987).

\bibitem{japan} A. Igarashi, J. Phys. Soc. Japan, {\bf 75}, 114301 (2006)

\bibitem{cffP} C. Froese Fischer and F. A. Parpia, Physics Letters A
{\bf 179}, 198 (1993).

\bibitem{japan2} A. Igarashi, J. Phys. Soc. Japan, {\bf 76}, 054301 (2007).

\bibitem{grant} I. P. Grant, J. Phys. B: At. Mol. Opt. Phys. {\bf 41}, 055002 
(2008).

\bibitem{dbsr} O. Zatsarinny and K. Bartschat, Phys. Rev. A {\bf 77}, 062701
(2008).

\bibitem{bsr} O. Zatsarinny, Comp. Phys. Commun. {\bf 174}, 273 (2006).

\bibitem{bachau} H. Bachau, E. Cormier, P. Decleva, J. E. Hansen, and F.
Martin, Rep. Prog. Phys. {\bf 64}, 1815 (2001).

\bibitem{deboor} C. de Boor, {\it A Practical Guide to Splines} (Spriner, 1978).

\bibitem{NG} P. H. Norrington and I.P. Grant, J. Phys. B, {\bf 14}, L261 (1981).

\bibitem{SH} R.Szmytkowski and J. Hinze, J. Phys. B, {\bf 29}, 761 (1996).

%\bibitem{grant08} I. P. Grant, J. Phys. B: {\bf 41} (2008).
%\bibitem{SO} S. Salomonson and P. \"Oster, Phys. Rev. A {\bf 40}, 5548 (1989).
%\bibitem{goldman} S. P. Goldman, Phys. Rev. A {\bf 31}, 3541. (1985).
%\bibitem{Slater} Y. Qiu and C. Froese Fischer, J. Comput. Phys.
%{\bf 156}, 257-271 (1999).


%\bibitem{morrison} J. Morrision, J. Phys. B: {\bf 29}, 2375 (1996).

%\bibitem{book} C. Froese Fischer, T. Brage, and P. J\"onsson, {\sl
%  Computational Atomic Structure: An MCHF Approach} (Inst. of Physics Publishing, Bristol 1997).


\end{thebibliography}
\end{document}